\documentclass[prd,a4paper,nofootinbib,showpacs,twocolumn]{revtex4}
\usepackage[a4paper, hdivide={1.9cm,,1.4cm}, vdivide={2.0cm,,3.3cm}]{geometry}

\usepackage{amstext,amssymb}
\usepackage[intlimits]{amsmath}
\usepackage[english]{babel}
\usepackage[ansinew]{inputenc}
\usepackage{graphicx}
\usepackage{slashed} 
\usepackage{units}
\usepackage{color}
\usepackage{hyperref}

\hypersetup{
    pdftitle={Sterile Neutrino Anarchy},
    pdfauthor={Julian Heeck, Werner Rodejohann}
}

\begin{document}

\let \Lold \L
\def \L {\mathcal{L}} 
\let \epsilonold \epsilon
\def \epsilon {\varepsilon} 
\let \arrowvec \vec
\def \vec#1{{\boldsymbol{#1}}}
\newcommand{\del}{\partial}
\newcommand{\dd}{\mathrm{d}}
\newcommand{\matrixx}[1]{\begin{pmatrix} #1 \end{pmatrix}} 
\newcommand{\tr}{\mathrm{tr}}
\newcommand{\hc}{\mathrm{h.c.}}
\newcommand{\diag}{\mathrm{diag}}
\def \M {\mathcal{M}} 

\renewcommand{\thefootnote}{\fnsymbol{footnote}} 

\title{Sterile Neutrino Anarchy}

\author{Julian \surname{Heeck}}
\email{julian.heeck@mpi-hd.mpg.de}
\affiliation{Max-Planck-Institut f\"ur Kernphysik, Saupfercheckweg 1, 69117 Heidelberg, Germany}

\author{Werner \surname{Rodejohann}}
\email{werner.rodejohann@mpi-hd.mpg.de}
\affiliation{Max-Planck-Institut f\"ur Kernphysik, Saupfercheckweg 1, 69117 Heidelberg, Germany}


\begin{abstract}
Lepton mixing, which requires physics beyond the Standard Model, is surprisingly compatible with a minimal, symmetryless and unbiased approach, called anarchy. This contrasts with highly involved flavor symmetry models. On the other hand, hints for light sterile neutrinos have emerged from a variety of independent experiments and observations. If confirmed, their existence would represent a groundbreaking discovery, calling for a theoretical interpretation. We discuss anarchy in the two-neutrino eV-scale seesaw framework. The distributions of mixing angles and masses according to anarchy are in agreement with global fits for the active and sterile neutrino parameters. Our minimal and economical scenario predicts the absence of neutrinoless double beta decay and one vanishing neutrino mass, and can therefore be tested in future experiments. 
\end{abstract}

\maketitle


\section{Introduction}

\addtocounter{footnote}{2} 

Neutrino oscillations have established the need for massive neutrinos and hence physics beyond 
the Standard Model (SM). While most of the data can be explained in a minimal framework of three 
active neutrinos---realized for example with a type-I seesaw mechanism~\cite{seesaw}---some 
experiments and observations strongly favor one or two additional light neutrinos, with masses around 
$\unit[1]{eV}$ and $\mathcal{O}(0.1)$ mixing~\cite{anomalies,Kopp:2011qd,Giunti:2011gz}. These hints stem from 
completely independent fields, namely particle physics, cosmology and astrophysics. We refer to the 
exhaustive overview of the situation provided by the White Paper in 
Ref.~\cite{whitepaper}. 
Since these new states are not allowed to contribute to the $Z$ width, they have been 
dubbed sterile neutrinos. A large number of experiments will be coming up in the next 
months and years with the aim of investigating their possible presence with a variety of methods and 
approaches~\cite{whitepaper}. 
Obviously, proving the existence of light sterile neutrinos 
would be a sensational and groundbreaking discovery. 

Sterile neutrinos aside, the mixing of active neutrinos has been of much interest to the physics 
community, as the peculiar values for the mixing angles seem to hint at an underlying symmetry 
principle. The astonishing precision of neutrino experiments has however put a slight 
dent in this idea, as increasingly complicated models seem to be required in order to be valid. On the 
other end of the model-building spectrum, the complete absence of a symmetry behind lepton 
mixing has also been proposed~\cite{anarchy}, both for an effective theory of neutrino mass and for the type-I seesaw case.  
The apparent randomness of parameters in this scheme can be the result of a sufficiently complex overlying fundamental theory, as discussed in detail in Ref.~\cite{Haba:2000be}.
This ``anarchy solution'' (henceforth called active anarchy) has been successful, especially in 
light of the rather large reactor mixing angle $\theta_{13}$~\cite{anarchy2}.

Taking the various hints for sterile neutrinos seriously, it is of obvious interest to try to 
accommodate these states in models. 
While flavor symmetry models that include sterile neutrinos exist~\cite{whitepaper}, 
the counterframework of anarchy has not yet been discussed in the context of light sterile 
neutrinos in detail. This will be amended by the paper at hand. 
Instead of simply adding two light sterile neutrinos to the three active ones 
(which would not be successful in the context of anarchy due to the sizable mass 
hierarchy between the active and sterile neutrinos), we will work in a 
much more economical scenario. Namely, we assume that the sterile 
states are the right-handed neutrinos 
of the type-I seesaw mechanism (eV-seesaw~\cite{lowenergyseesaw}). 
The most minimal case which can accommodate the active neutrino data is then 
when only two such states are added to the picture.\footnote{
We note further that anarchy with $n$ (heavy seesaw) right-handed neutrinos has been recently 
discussed in Ref.~\cite{Heeck:2012fw}, where it has been found that the active neutrino 
mass hierarchy prefers small $n$.} 
This implies two light sterile neutrinos, 
one massless active neutrino, no neutrinoless double beta decay and 
only a small effective electron neutrino mass. A 
phenomenological study of this straightforward scenario has recently been 
provided in Ref.~\cite{Donini:2011jh}. 
We will study the implications of anarchy in this framework and present the 
statistical distributions for all observables. We show that the large 
active--sterile mixing angles needed for various anomalies can be naturally obtained. 
Thus, the economic and attractive scenario of anarchy can also be extended to 
the case of light sterile neutrinos---in particular the minimal two-neutrino eV-scale seesaw.

\section{Anarchy}

Two right-handed singlet neutrinos $N_{1,2}$ modify the SM Lagrangian by the following terms:
\begin{align}
\begin{split}
 \L \ &= \ \L_\mathrm{SM} +i \overline{N}_j \slashed{\del} N_j -\left( \overline{N}_j  \left( \vec{Y}_\nu \right)_{j \alpha} \tilde{H}^\dagger L_\alpha \right.\\
&\quad \left. + \tfrac{1}{2} \,\overline{N}_j \left( \M_R \right)_{jk} N_k^c + \hc \right),
\end{split}
\end{align}
where sums over $j,k=1,2$ and $\alpha = e,\mu,\tau$ are understood.
After electroweak symmetry breaking, using the usual SM Higgs doublet $H\rightarrow (0,v)^T \simeq (0,\unit[174]{GeV})^T$, we arrive at the $5\times 5$ Majorana mass matrix for the neutral fermions in the basis $(\nu_L, N^c)^T$:
\begin{align}
\begin{split}
 \M_\mathrm{full} &= \matrixx{ 0 & v V_L^* D_Y^T V_R^T \\ v V_R D_Y V_L^\dagger & \M_R} \\
&= U^* \,\diag (m_1, m_2, m_3, m_4,m_5)\, U^\dagger \,.
\label{eq:mfull}
\end{split}
\end{align}
Here, we used the singular value decomposition for the Yukawa coupling matrix
\begin{align}
 \vec{Y}_\nu = V_R \matrixx{y_1 & 0 & 0\\ 0 & y_2 & 0} V_L^\dagger\equiv V_R D_Y V_L^\dagger
\end{align}
and introduced a unitary matrix $U$ to diagonalize $\M_\mathrm{full}$.

In the anarchy framework we assume that the unitary $3\times 3$ ($2\times 2$) matrix $V_{L}$ ($V_R$) is distributed according to the Haar measure of $U(3)$ ($U(2)$), which can then be compared to the experimental values.\footnote{Note that it does not matter whether we take a diagonal $\M_R = \diag (M_1,M_2)$ or $V \M_R V^T$ with Haar-distributed $V\in U(2)$, because our framework is by construction basis independent.} As far as the singular values $y_j$ and eigenvalues of $\M_R$ go, we assume a distribution according to the linear measures as derived in Refs.~\cite{Haba:2000be,Heeck:2012fw}:
\begin{align}
\begin{split}
 \dd \M_R &\propto |M_1^2 - M_2^2| M_1 M_2 \,\dd M_1 \dd M_2\,, \\
 \dd D_Y &\propto (y_1^2 - y_2^2)^2 y_1^3 y_2^3 \,\dd y_1 \dd y_2\,, 
\end{split}
\end{align}
with the boundary conditions $\tr (D_Y^\dagger D_Y) = \sum_i y_i^2 \leq y_0^2$ and $\tr (\mathcal{M}_R^\dagger \mathcal{M}_R ) = \sum_i M_i^2 \leq M_0^2$. A survey of more complicated measures including weighting functions, analogous to Ref.~\cite{Jenkins:2008ms} for active anarchy, lies beyond the scope of this paper.

Note that this is the proper way of defining anarchy in the case of two right-handed neutrinos, even outside of the seesaw limit $\M_R \gg v \vec{Y}_\nu$. Taking for example the full mass matrix $\M_\mathrm{full}$ to be distributed according to $U(5)$ (as in Ref.~\cite{Gluza:2011nm}) would make Majorana masses for the $\nu_L$ necessary, for which we would have to introduce a scalar triplet (type-II seesaw), complicating the study considerably. Additional mechanisms that decouple some heavy right-handed neutrinos or impose anarchy in only a subsector could also be constructed~\cite{modifiedanarchy}, but we will only consider the minimal framework here.

\begin{figure}[tb]
\setlength{\abovecaptionskip}{-2ex}
	\begin{center}
		\includegraphics[width=0.43\textwidth]{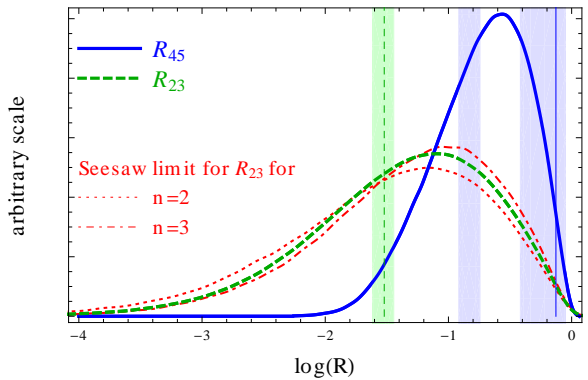}
	\end{center}
		\caption{Distribution of mass ratios $R_{ij}\equiv \Delta m_{i1}^2/\Delta m_{j1}^2$ for the values $M_0 = \unit[1.6]{eV}$, $y_0 = 1.5\times 10^{-12}$. The red/dotted (red/dot-dashed) distribution corresponds to $R_{23}$ in the seesaw limit with two (three) sterile neutrinos. Vertical lines denote best-fit values, shaded areas $90\%$~C.L.~(sterile) and $3\sigma$~(active) ranges.}
	\label{fig:ratios}
\end{figure}

In Eq.~\eqref{eq:mfull}, we have $m_1=0$, and the only free parameters are the scales $y_0$ and $M_0$. $M_0$ can be fixed to give $m_5\simeq \unit[1]{eV}$, while the Yukawa scale $y_0$ is fixed to render $m_3\simeq\unit[0.05]{eV}$.\footnote{A small scale $M_0$ is technically natural in that $M_0 \rightarrow 0$ enhances the symmetry of the Lagrangian~\cite{Fujikawa:2004jy,lowenergyseesaw}.} The other two masses and all the mixing angles and phases are then distributed in a known way and can be compared to measurements. Defining the parameter $\epsilon \equiv v y_0/M_0$, we can estimate the expected values of the observables:
\begin{align}
\begin{split} \nonumber 
 &m_{2,3} \sim M_0 \,\epsilon^2 \,( 1+ \mathcal{O}(\epsilon^2) ) \,, \quad\nonumber 
 m_{4,5} \sim M_0\, ( 1+ \mathcal{O}(\epsilon^2) ) \,, \\ \nonumber 
 &U_{\alpha 4},U_{\alpha 5} \sim \epsilon \,( 1+ \mathcal{O}(\epsilon^2) ) \,, \quad
 U_\mathrm{PMNS} \sim ( 1+ \mathcal{O}(\epsilon^2) ) \,.\nonumber 
\end{split}
\end{align}
Here, $U_\mathrm{PMNS}$ denotes the upper-left $3\times 3$ submatrix of $U$ (corresponding to the Pontecorvo--Maki--Nakagawa--Sakata (PMNS) mixing matrix), which contains the active mixing angles. 
In the eV-seesaw limit that we are interested in, $\epsilon \lesssim 0.2$. The case 
$\epsilon \rightarrow 0$ is the canonical seesaw limit. 
We can estimate that in case of the eV-seesaw limit, the ratios $ R_{23}\equiv \Delta m_{21}^2/\Delta m_{31}^2$ and $ R_{45}\equiv \Delta m_{41}^2/\Delta m_{51}^2$  will only receive percent corrections compared to the seesaw limit. This can be seen in Fig.~\ref{fig:ratios}, where the distributions for $R_{23}$ are shown for $\epsilon \simeq 0.2$, $\epsilon \ll 0.2$ ($n=2$) and $\epsilon \ll 0.2$ ($n=3$):
the hierarchy of $R_{23}(n=2)$ is pulled closer to $R_{23}(n=3)$ due to the next-to-leading order seesaw contributions.\footnote{Here and in the following, all shown distributions are properly normalized. 
The sampling procedure follows Ref.~\cite{Heeck:2012fw}.}

\begin{figure}[tb]
\setlength{\abovecaptionskip}{-2ex}
	\begin{center}
		\includegraphics[width=0.43\textwidth]{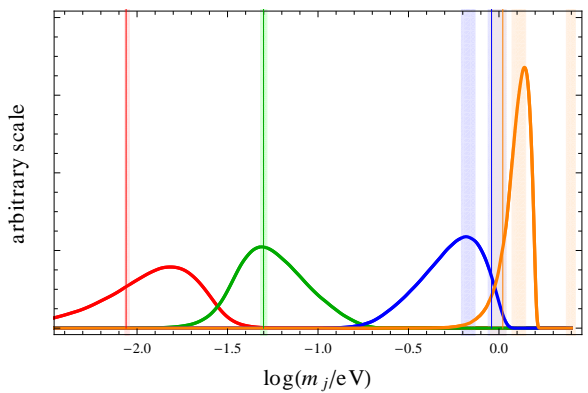}
	\end{center}
		\caption{Distribution of masses for the same values as in Fig.~\ref{fig:ratios}.}
	\label{fig:masses}
\end{figure}

\begin{figure}[tb]
\setlength{\abovecaptionskip}{-2ex}
	\begin{center}
		\includegraphics[width=0.43\textwidth]{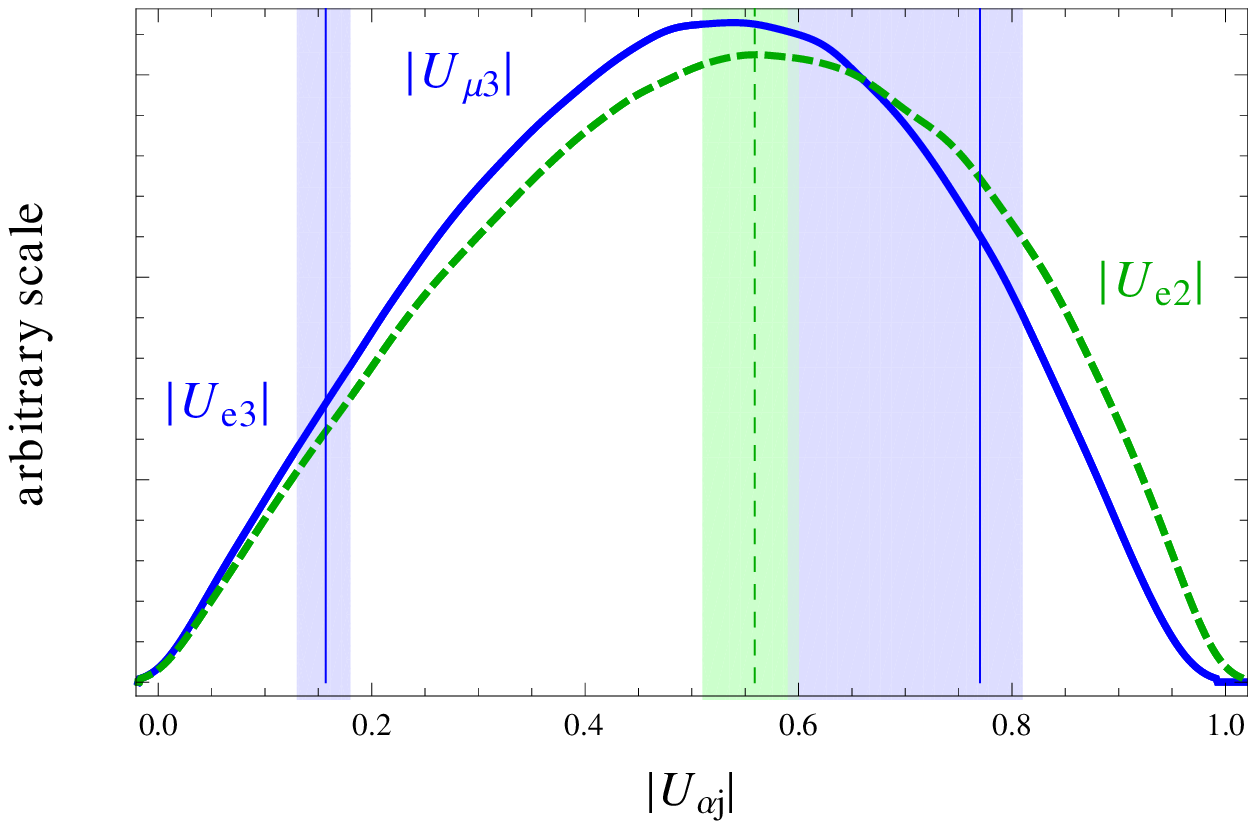} \\
		\includegraphics[width=0.43\textwidth]{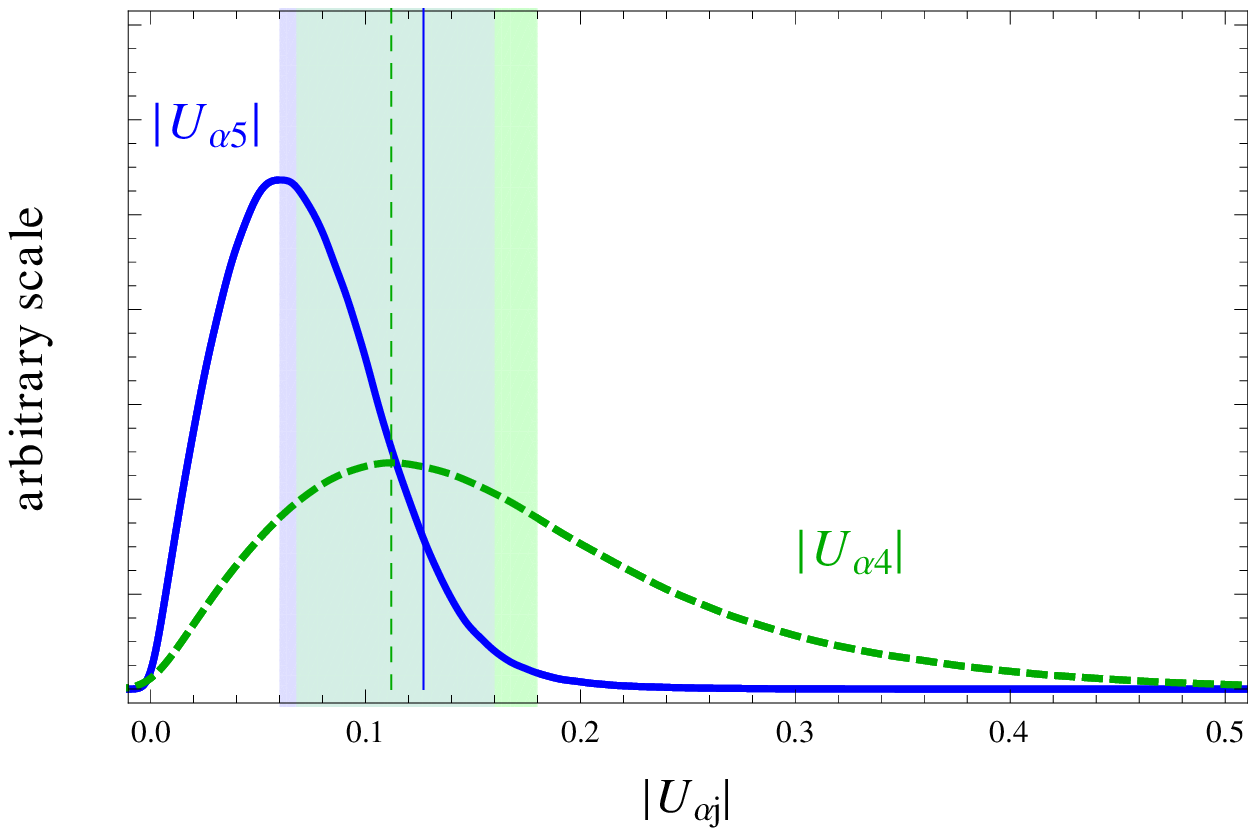}
	\end{center}
		\caption{Distribution of some mixing elements for the values $M_0 = \unit[1.6]{eV}$, $y_0 = 1.5\times 10^{-12}$. The vertical lines represent best-fit values and the shaded areas are $3\sigma$ ranges (top from Ref.~\cite{Tortola:2012te}, bottom for $\alpha = \mu$ from Ref.~\cite{minimal3plus2}). In the seesaw limit, $U_{\alpha 5}$ and $U_{\alpha 4}$ go to zero, while $U_{\alpha 3}$ and $U_{\alpha 2}$  converge roughly to the distribution for $U_{e 2}$ (top).}
	\label{fig:mixing}
\end{figure}

\begin{figure}[tb]
\setlength{\abovecaptionskip}{-2ex}
	\begin{center}
		\includegraphics[width=0.43\textwidth]{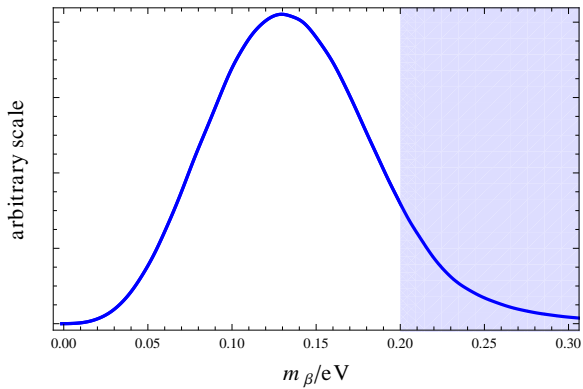}
	\end{center}
		\caption{Distribution of $m_\beta$ for the values $M_0 = \unit[1.6]{eV}$, $y_0 = 1.5\times 10^{-12}$. The shaded region corresponds to the $90\%$~C.L.~design sensitivity of KATRIN~\cite{katrin}.}
	\label{fig:katrin}
\end{figure}

As interesting numerical values throughout this paper we use $M_0 = \unit[1.6]{eV}$, $y_0 = 1.5\times 10^{-12}$ (i.e.~$\epsilon \simeq 0.16$). In our Figs.~\ref{fig:ratios}--\ref{fig:katrin}, we show the distributions of some observable quantities as well as current global-fit values. The best-fit values and $3\sigma$ ranges for the active neutrino parameters are taken from Ref.~\cite{Tortola:2012te}; the values for the active--sterile mixing elements $|U_{\alpha 4,5}|$ from Ref.~\cite{minimal3plus2}; and the masses $\Delta m^2_{41}$, $\Delta m^2_{51}$ from Ref.~\cite{whitepaper} (Fig.~71). Using other global fits for these parameters, e.g.~from Refs.~\cite{Kopp:2011qd,Giunti:2011gz}, does of course not change the qualitative discussion of this paper. We further note that the precision in the parameters associated with the sterile neutrinos is much weaker than it is for the active ones.

Let us emphasize that while the best-fit value for $R_{45}$ is somewhat larger than the anarchy prediction, this tension is alleviated once we consider e.g.~the $90\%$~C.L.~contour. For example, there seems to be an interesting region in parameter space $\Delta m_{41}^2 \simeq \unit[1]{eV^2}$, $\Delta m_{51}^2 \simeq \unit[6]{eV^2}$~\cite{whitepaper} that gives roughly $R_{45} = 0.17$, perfect for sterile anarchy. Obviously, the agreement with standard $\Lambda$CDM cosmology worsens in this case due to the large sum of neutrino masses~\cite{sterilecosmology}. Note also that the preference of two sterile neutrinos over one has weakened~\cite{giunti} due to a recent update from the MiniBooNE experiment~\cite{miniboone}, which reduces the previous tension between the $\nu_\mu\rightarrow \nu_e$ and $\overline{\nu}_\mu\rightarrow \overline{\nu}_e$ data. Furthermore, a stringent $99\%$~C.L.~limit for the active--sterile mixing angle of $\sin\theta <0.07$ in the relevant mass range was given very recently by the ICARUS experiment~\cite{icarus}. However, from Fig.~\ref{fig:mixing} (bottom) we see that a smaller mixing angle is also fine in sterile anarchy.

Let us discuss the distribution of the mixing matrix $U$. Since we are not working in the seesaw limit, or in an effective theory, the upper-left $3\times 3$ submatrix $U_\mathrm{PMNS}$ is not distributed according to the Haar measure of $U(3)$ as in active anarchy~\cite{Haba:2000be}
\begin{align}
 \dd U_\mathrm{PMNS} \propto \dd s_{12}^2\, \dd s_{23}^2\, \dd c_{13}^4\, \dd \delta\, \dd \alpha\, \dd \beta
\end{align}
(which would imply the same distribution for all $|U_{\alpha j}|$, $\alpha = e,\mu,\tau$, $j=1,2,3$), but shows small deviations. From the unitarity of $U$---namely $\sum_i |U_{\alpha i}|^2 = 1$---we expect diminished values for $|U_{\alpha 1,2,3}|$ upon increasing the active--sterile mixing $|U_{\alpha 4,5}|$. The diagonalization condition $\sum_i U_{\alpha i}^2 m_i = 0$---related to the upper-left $3\times 3$ zero matrix in $\M_\mathrm{full}$---then shows that mainly $|U_{\alpha 3}|$ is suppressed (due to $m_3>m_2$). 
This is illustrated in Fig.~\ref{fig:mixing} (top), where we see that $|U_{e 2}|$ (and therefore $\theta_{12}$) is still approximately distributed like in the seesaw limit, while $|U_{\alpha 3}|$ is drawn to slightly smaller values. This helps the agreement of anarchy with the rather small (in comparison to $\theta_{12,23}$) reactor mixing angle $|U_{e 3}|\simeq s_{13} \simeq 0.16$, but makes the agreement with the (comparably imprecise) atmospheric mixing angle $\theta_{23}$ a little worse---both, however, insignificant.

The active--sterile mixing elements $|U_{\alpha 4,5}|$ are large, as expected from the scaling $|U_{\alpha 4,5}| \sim \epsilon \sim 0.1$ (bottom Fig.~\ref{fig:mixing}). The suppression of $|U_{\alpha 5}|$ compared to $|U_{\alpha 4}|$ is due to the mass hierarchy in $\M_R$, as the entries scale with $1/\sqrt{m_5}$ and $1/\sqrt{m_4}$, respectively~\cite{lowenergyseesaw}. We are led to conclude that 
sterile anarchy inherits the success of active anarchy for the PMNS mixing matrix, and is further fully compatible with large active--sterile mixing angles.

Having discussed the distributions of masses and mixing angles, we now briefly turn to other 
observables. First of all, as in any anarchy scenario, the normal mass ordering is preferred. 
In our case, only about 5\,\% of the cases give the inverted ordering. 
The relevant phase for short-baseline experiments 
$\arg (U_{e 4}^* U_{e 5} U_{\mu 4} U_{\mu 5}^* )$ is distributed 
uniformly from zero to $2 \pi$---as expected from phases in anarchy---and can therefore not 
be used to test this framework.
The rate of neutrinoless double beta decay~\cite{betadecay} vanishes; this is because all 
neutrino masses are far below the momentum exchange $q^2\simeq (\unit[100]{MeV})^2$ 
relevant for the decay, so the matrix element will be proportional to $(\M_\mathrm{full})_{ee}=0$.
Kurie-plot (beta decay) experiments can test the effective electron neutrino mass
\begin{align}
 m_\beta \equiv \sqrt{\sum\nolimits_j |U_{e j}|^2 m_j^2}
\end{align}
with a current upper limit of $\unit[2.3]{eV}$ at $95\%$~C.L.~\cite{betalimit}. 
The distribution of $m_\beta$ is shown in Fig.~\ref{fig:katrin}. It 
is expected to be rather small; hence, a discovery in the KATRIN experiment~\cite{katrin} (detection potential $m_\beta = \unit[0.35]{eV}$) would pretty much exclude sterile anarchy.

\section{Conclusion}
Anarchy for active neutrinos both with and without a seesaw mechanism has been shown to be in 
good agreement with the measured neutrino mixing parameters, and is much more economical than 
typically studied flavor symmetries. 
In this paper we have demonstrated that an eV-scale type-I seesaw mechanism with two 
right-handed neutrinos can extend this framework to a minimal scheme with light sterile 
neutrinos. The mass hierarchies and mixing angles are surprisingly close to the ones 
needed to explain a number of neutrino anomalies. Statistical 
improvements for the sterile neutrino parameters are necessary to properly evaluate 
the validity of the approach discussed here. While currently not really predictive 
due to the low precision of sterile neutrino parameters, the model can be easily 
excluded, as it hinges on the absence of neutrinoless double beta decay, no discovery 
at KATRIN and the existence of two light sterile neutrinos, all of which are falsifiable 
in the near future. This is to be compared to active anarchy, 
which is notoriously hard to test for. Nevertheless, we have presented here the most minimal and 
economic framework to explain active and sterile neutrino parameters.  

\vspace{-.2cm}
\begin{acknowledgments}
This work was supported by the Max Planck Society through the Strategic Innovation Fund 
in the project MANITOP. J.H.~acknowledges support by the IMPRS-PTFS.
\end{acknowledgments}

\end{document}